# Entropy-based Prediction of Network Protocols in the Forensic Analysis of DNS Tunnels

Irvin Homem, Panagiotis Papapetrou

Department of Computer and Systems Sciences
Stockholm University
Sweden
{*irvin, panagiotis}@dsv.su.se

Spyridon Dosis

KTH-Royal Institute of Technology
Sweden
dossis@kth.se

*Abstract*—DNS tunneling techniques are often used for malicious purposes but network security mechanisms have struggled to detect these. Network forensic analysis has thus been used but has proved slow and effort intensive as Network Forensics Analysis Tools struggle to deal with undocumented or new network tunneling techniques. In this paper we present a method to aid forensic analysis through automating the inference of protocols tunneled within DNS tunneling techniques. We analyze the internal packet structure of DNS tunneling techniques and characterize the information entropy of different network protocols and their DNS tunneled equivalents. From this, we present our protocol prediction method that uses entropy distribution averaging. Finally we apply our method on a dataset to measure its performance and show that it has a prediction accuracy of 75%. Our method also preserves privacy as it does not parse the actual tunneled content, rather it only calculates the information entropy.

## I. INTRODUCTION

In recent years there has been an increase in the use of DNS tunneling to stealthily perpetrate malicious activity. This is often in the form of exfiltration of sensitive data, hiding of network attacks and the perpetuation of malware activities through botnet communication [1]. Several strains of malware such as the Morto worm [2], Feederbot [3] and variants of POS (Point Of Sale) malware such as the BernhardPOS and the FrameworkPOS malware [4] show the rise in popularity of this method of stealthy communication. The availability of several freely available tools for performing DNS tunneling (such as NSTX, Iodine, dnscat, DeNiSe OzymanDNS and Heyoka) [5] have also aided its popularity and uptake.

Preventive measures fail at curtailing such breaches [4]. DNS tunneling detection mechanisms have been developed [6] [7], however they also struggle, discovering only 3% of attacks in sophisticated real world cases [4]. Work has been done to improve the detection of tunneling as is seen in [8] and [9], however despite these efforts, network breaches involving tunneling are still on the increase [4] [2].

Reactive security mechanisms such as network security monitoring and network forensics analysis techniques offer some promise, however the process is often manual and effort intensive [10]. It also requires highly skilled expertise and is time-consuming, taking up to 7 months [4] [11]. Furthermore, current industry standard network forensics analysis tools only parse standardized network protocols. Previously unseen or undocumented network protocols common with DNS tunneling require manual dissection [10]. New and innovative methods to help alleviate these challenges in order to speed up the forensic analysis of tunneled network traffic are thus needed.

In the network forensic analysis of tunneled networked traffic, it is of primary concern to identify the carrier tunneling protocol, the internally tunneled protocol, the communicating parties, the content that is being tunneled and its significance. Within our focus on DNS tunneling techniques, we assume that the identification of DNS tunneling has already been done, as there are studies such as [6] and [9] that demonstrate such methods. Thus we shift our focus to the next important process which is the discovery of the network protocol that is being carried internally within the DNS tunnel.

To the best of our knowledge, there is an absence of studies done on the forensic analysis of DNS tunneling techniques. There has also been no work done on the identification of network protocols being tunneled within DNS tunnels. We thus focus on this area, developing a prediction mechanism to probabilistically identify network protocols carried within DNS tunnels. To this end, we hypothesize that individual network protocols inherently exhibit a specific entropy distribution in their byte contents in their normal usage. We further postulate that network protocols carried within tunneling mechanisms maintain some similarity to their original byte entropy distributions. This enables us to probabilistically match DNS tunneled traffic with reasonable accuracy to particular protocols matching the same distribution.

As a starting point, we limit our study to the identification of two network protocols: HTTP and FTP, as tunneled individually by a single DNS tunneling tool. We also do not concern ourselves, for now, with the identification of the IP addresses of the communicating parties, nor the exact content being transmitted within the messages of the internally tunneled protocol. These are to be considered as future work. Additionally, for the sake of simplicity we focus on the popular Iodine DNS tunneling tool, though we believe that our idea can be similarly applied to other tools using different DNS tunneling methods. For the moment we do not deal with multi-level nesting of tunneled protocols. We also do not deal with encrypted protocols carried within the tunneling protocol, nor tunneling protocols that inherently use encryption such as IPSec, SSH or SSL. The reason for this is that encryption methods aim at making information entropy uniform, per block or stream, such that different parts of the message are indistinguishable from each other. However, these are to be considered as future work.

A network protocol prediction tool is built based on our theory. Initial outcomes from this tool show promising results with a prediction accuracy of 75% indicating a positive sentiment towards the development of our theory. Given the potential for large volumes of network traffic captures in today's digital investigations, our protocol prediction tool could help speed up the triage process as well as the analysis of network traffic by helping forensic analysts to identify particular network flows where a certain suspect protocol may be present within network traffic, but is hidden in DNS tunneling techniques.

## II. BACKGROUND AND RELATED WORK

Few studies have been performed to determine the protocols that are being carried within a tunneling protocol. In [12] traffic classification of protocols tunneled over SSL was performed using only the sizes of the first few packets of a given SSL session. They first identify SSL traffic from other normal traffic and then use a clustering mechanism based on Gaussian mixture models to distinguish between several protocols (including HTTP, FTP, BitTorrent, edonkey, SMTP, and POP3) tunneled over SSL. Statistical analysis of packet sizes and inter-arrival times were used in [13] to fingerprint normal SSH usage, and when SSH is used for tunneling other protocols. They extended their work in [14] to also distinguish HTTP tunneling from normal HTTP traffic, as well as to predict the presence of plaintext protocols such as POP3, SMTP, Chat and P2P protocols within both SSH and HTTP tunnels. Adaboost, C4.5 and Genetic Programming based classifiers were used in [15] to distinguish Skype and SSH traffic from other traffic, as well as to identify the type of application traffic (Shell, SFTP, SCP, Local /Remote Forwarding or X11) being tunneled in the SSH tunnel.

Other notable studies have focused on identifying proprietary protocols such as Skype and Spotify among other network traffic. These studies include [16], [17] and [18]. In terms of content inference in tunnels, [19] attempted to extract exact key strokes from encrypted live SSH shell sessions.

With regard to DNS tunneling techniques, most studies have focused only on detection. In [6] an n-gram character frequency analysis method for identifying domain names typical of DNS tunneling traffic was described. In [1] an anomaly detection method based on contrasting the statistical and information theoretic properties of the payload content of normal DNS traffic against that of DNS tunneling techniques was presented. Several DNS tunneling tools and detection heuristics are discussed in [9] including DNS request and response sizes, domain name entropy, use of uncommon Resource Records, the volume of DNS requests per IP address or per domain, the number of subdomains per domain, and presence of large numbers of orphaned DNS requests.

Initial methods for *manual* disassembly of DNS tunneling traffic to recover the internally carried protocols and data are seen in [10]. To the best of our knowledge there are no studies on the automated prediction of network protocols tunneled within DNS tunneling traffic for the benefit of forensic analysis, thus we direct our efforts in this study towards this.

## III. DNS TUNNEL INTERNALS AND DATASET COLLECTION

The flexibility of the DNS protocol allows DNS tunneling tools to use varying techniques [20]. Many tools append data as a subdomain in the name field of queries, though they also each have variations aimed at ease of use, increase in throughput or increased invisibility to security mechanisms. For example the "dns2tcp" tool uses TXT records; Iodine uses NULL records, while DNScat uses CNAME records. Iodine and Heyoka use EDNS(0) extensions in order to increase throughput [20].

### A. Tunneling with IoD-ine (IP-Over-DNS)

In this study we use the Iodine DNS tunneling tool due to its popularity, ease of use and availability of documentation [10]. Iodine encapsulates IPv4 packets into the payload of DNS packets. It uses NULL resource records by default, but can also use other resource records such as PRIVATE, TXT, SRV, MX, CNAME and A. The Query/Answer *name* field, within the particular Resource Record in use, holds the encapsulated data.

Upstream data is GZIP compressed and encoded. Encoding options include Base32, Base64 (or Base64u) and Base128. This is determined through checking for character set support at intermediate DNS servers. Downstream data is transmitted as GZIP compressed raw IP packet bytes, if using NULL or PRIVATE resource records. If other resource records are used, then downstream data is GZIP compressed *and* encoded like upstream data [21]. If encoding is applied, then the downstream header is prepended with a specific ASCII character signifying the encoding type used.

Tunneled data within the *name* field consists of a header and the fragment of the packet being tunneled, prepended as a sub-domain of the tunneling server. The header preceding the tunneled data contains metadata such as the user id, the codecs in use (Base32/64/128), the fragment size, the fragment number, the sequence number, whether compression is used and a Cache Miss Counter [21].

### B. DNS Tunneling Setup and Dataset Capture

As there are no well-known DNS tunneling network traffic captures available and it is difficult to obtain network traffic captures involving malicious DNS activity [22] we set up a DNS tunnel using the Iodine tool to create our own dataset.

To create our DNS tunneled traffic dataset, we simulated the use of HTTP and FTP protocols individually, each within its own DNS tunnel. HTTP traffic over DNS was generated by performing simple web requests to 8 websites allowing for any additional requests for extra content (images, CSS, JavaScript, Ads) to ensue. For simulating the use of the FTP protocol we downloaded several files that we had placed on an FTP server prior to the experiments. In order to include some variation we ensured that the files downloaded were of different types including images, PDFs, text files, audio, video and ZIP files. The files downloaded were 12 in total and were stored at different paths on the FTP server. Summarily the dataset included 20 DNS tunneled traffic samples. 8 were of HTTP communication, each with one of the 8 specific websites. The other 12 were of FTP communication involving logins, directory traversals and individual file downloads.

As a control dataset for comparison of the tunneled network protocols against their plain versions, we simulated normal HTTP web traffic and normal FTP traffic. The HTTP traffic was generated by visiting a randomly chosen website with a lot of content to be loaded. To generate normal FTP traffic we performed a login to an FTP server and sequentially downloaded 3 different files. We chose to use multiple files due to the terse nature of the FTP protocol commands.

## IV. PROTOCOL FEATURE TRENDS & ANALYSIS

Our method for network protocol prediction is based on identifying patterns based on features found both in the plain protocol traffic that can be mapped to equivalent features in the DNS tunneled network traffic. Several features could be chosen to characterize the differences between protocols, and similarities across the plain and tunneled versions, including byte frequencies, information entropy and packet lengths. We chose to limit this study to the analysis of information entropy, as a starting point, as it is inherently tied to the actual data bytes composing packets. The idea is to observe normal traffic of particular protocols (HTTP and FTP in our case) and to make comparisons with their DNS tunneled equivalents that have been fragmented, encoded and compressed in the tunneling process.

### A. Experimenting with Information Entropy

In characterizing features, measurements may be made at different levels of abstraction: At the IP packet level, the transport level or the application level. Differentiation may also be made between protocol client requests and server responses. Both HTTP and FTP protocols have a relatively small vocabulary of commands and content that goes into the *requests*; while their responses could include large amounts of data with great variation. We thus focus only on the requests of these protocols where the content and variation is more predictable and likely more significant for comparisons.

*Information entropy* is a measure of the variation of the components that make up a message. In our case, we compute this on the bytes that make up a packet layer or field value. Entropy, as applied here, is calculated as the probability of a particular byte occurrence $p(x_i)$ multiplied by the logarithm of the probability of that occurrence, summed up for all byte occurrences. The mathematical formula is depicted below:

$$H(X) = -\sum_{i=1}^{n} p(x_i) \log p(x_i)$$

The hypothesis behind the use of entropy is that the request packets for a particular protocol flow will produce a specific entropy distribution. We test this by creating a simple Python program (using the Scapy and Matplotlib libraries) to measure the information entropy feature and plot some charts for visual analysis of the distribution trends as seen in Fig. 1 and Fig. 2

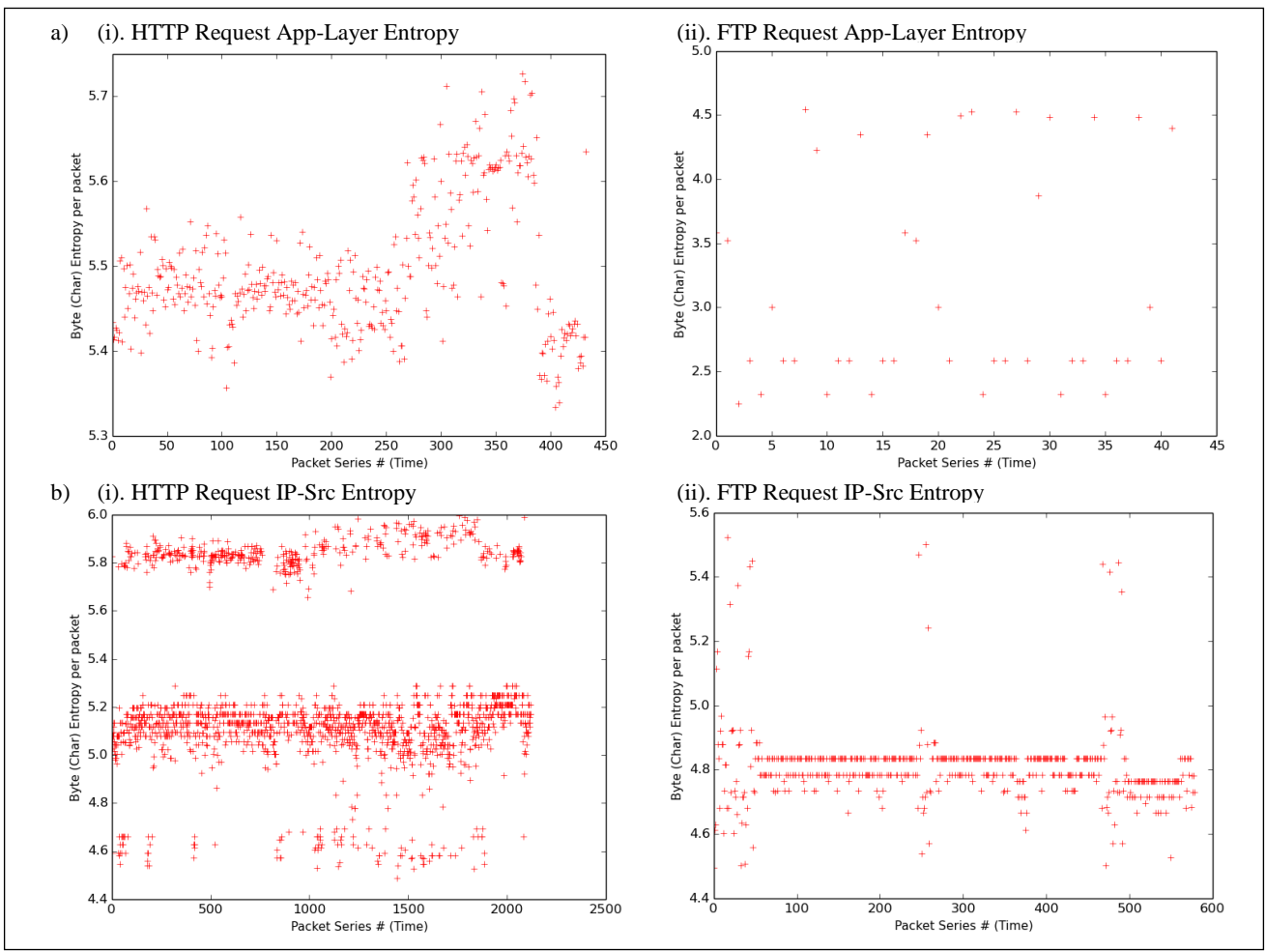

Figure 1. Comparison of Packet Entropy at Different Abstraction levels for Different Protocols

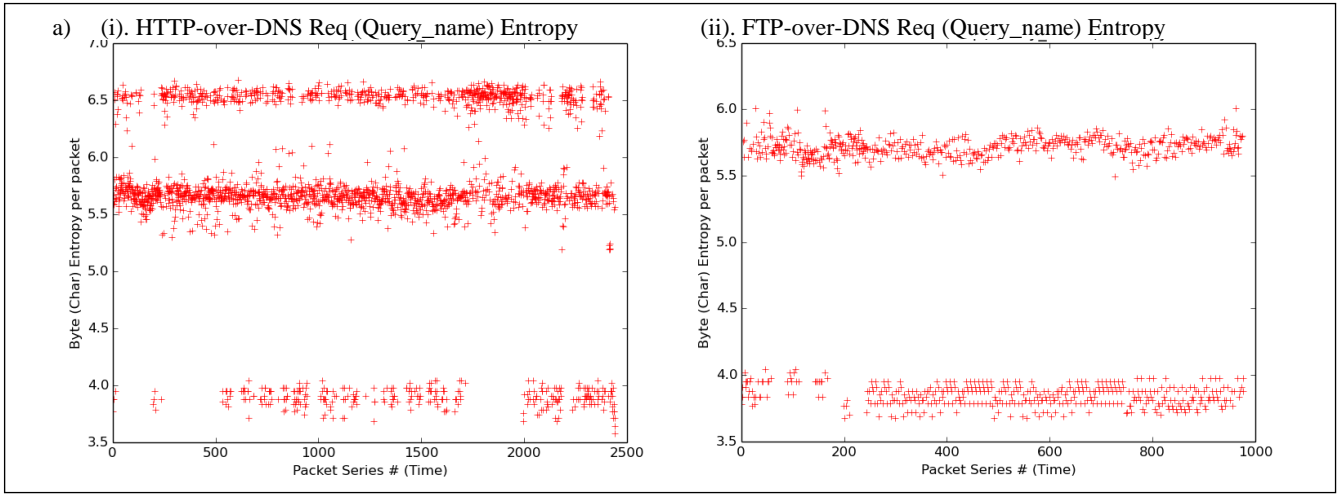

Figure 2. Comparison of Packet Entropy Values among Tunneled Protocols

## B. Comparison of Plain Network Protocols with Tunneled Equivalents

Fig. 1(a)(i) and (ii) show the distributions of the entropy of the *application layer requests* of plain HTTP and FTP network traffic respectively. This is filtered through taking only the packets that contain application layer content and are destined to port 80 and port 21, for HTTP and FTP respectively. Thus there are no transport layer features present and thus no ACKs seen as these ACKs do not have any application layer content.

Fig. 1(b)(i) and (ii) show entropy distributions at the *IP packet level* for plain HTTP and FTP traffic respectively. Filtering is performed at a more granular level where all IP traffic from the client is captured. The packet entropy distributions here contain both packets that contain application data and packets containing transport layer ACKs since both HTTP and FTP use TCP for the transport layer. These client ACKs acknowledge the receipt of prior Server S-ACKs that originated from the delivery of a request to the server

Fig. 2(a)(i) and (ii) show the entropy distributions of HTTP over DNS traffic and FTP over DNS traffic respectively. They show the entropy of the Query Name field for DNS requests destined for the DNS port 53. As the filtering is based on the client side traffic destined for port 53, it includes both the tunneled application protocol request packets as well as those embodying transport layer properties, e.g. ACKs, sequencing and reliability information. Filtering is performed this way as there are no established methods for parsing and differentiating application layer data and other layers for DNS tunneled traffic. Moreover it also preserves privacy, as only calculations are performed over the packet content to generate metadata (*entropy* in this case), without necessarily making sense of the actual content.

Given Base128 encoded traffic, which is the default used in Iodine, the theoretical maximum entropy for totally random data tends towards 8-bits. From the distributions in Fig. 1 and Fig. 2, there are variations seen between the protocols as well as between different layers of abstraction. The distributions are not uniformly distributed at the 8-bit value, evidencing an absence of absolute randomness. They also portray different patterns that may indicate the presence of a particular network protocol.

In Fig. 1(a)(i) and (ii) the HTTP application layer traffic entropy values are seen to be significantly denser, than that of FTP traffic. The FTP traffic seems to have application layer packet entropy values around 4.5 and others around 2.5. HTTP application layer traffic seems to have most of its entropy values between 5.4 and 5.6.

Similarly, in Fig. 1(b)(i) and (ii) the density of packets seems to be less with FTP traffic than with HTTP traffic. There seem to be 3 bands of clusters within the HTTP traffic, with the densest band being around the 4.9 to 5.3 entropy values. The other 2 bands are seen to be at around 4.6 and 5.8. With the FTP traffic there seems to be a single dense cluster of entropy values around the 4.8 mark. However one could also argue for 2 more highly sparse clusters around the 5.4 and 4.5 entropy values. These sparse clusters may be significant when one looks across on the sequence of packets where we hypothesize the variances in entropy away from the 4.8 mark indicate the FTP requests initiating downloads for the 3 different files.

With regard to the entropy values of HTTP over DNS traffic and FTP over DNS traffic depicted in Fig. 2(a)(i) and (ii), respectively, there is also some clustering. In the HTTP traffic, there seem to be 3 cluster bands. The densest is around 5.6-5.8, the next less dense band is around the 6.5 value and the least dense seems to be around the 3.8-4.0 entropy range. For the FTP over DNS traffic seen in Fig. 2(a)(ii) there are 2 main clusters of entropy values. One band around the 5.6-5.8 range of entropy values and the other around the 3.8-4.0 range.

Shorter message sizes or smaller alphabets result in smaller entropy values [3]. In the case of tunneled traffic with a fixed encoding of Base128, two types of traffic could consist of smaller entropy values due to shorter packet lengths. One is the ACKs from the transport layer. The second is the "ping" functionality often built into DNS tunneling tools to prevent DNS servers from timing out, by sending short query request messages periodically.

We postulate that the bands in both Fig. 2(a)(i) and (ii) around the 4.0 value correspond to the entropy of the pings as they are shorter (less than 30 bytes) than ACKs (at least 40 bytes) when seen in manual DNS tunneling disassembly. This is reinforced in that they are less dense in the HTTP traffic series due to the fact that a HTTP request can spawn several other HTTP requests to retrieve more content for the proper loading of a website. Thus the multitudes of HTTP requests keep the connection between the DNS tunneling client and the tunneling server open, corresponding to a lesser need for the pings to be generated for time-out prevention.

The bands around 5.6-5.8 in both the tunneled HTTP traffic and FTP traffic seen in Fig. 2(a)(i) and (ii), respectively are likely ACKs from the transport layer. This deduction is made mainly from the HTTP over DNS traffic containing another band with higher entropy, which is likely the HTTP requests which are the longest resulting in higher entropy values. Also, the small repetitive pattern seen in the first 200 packets in Fig. 2(a)(ii) seems to correspond to the action of the 3 file downloads. This leads us to believe that the band around the 4.0 mark contains also the FTP request packets hidden among the same cluster with the "pings". This is reinforced by the fact that the FTP commands making up the FTP requests are terse and have a fixed structure. This would contribute to a generally smaller entropy than HTTP requests that have a larger request header set and many more fields.

Due to this clustering of different types of packets from the effect that their content and lengths have on the entropy, we deduce that there are inherent distributions that can help distinguish between DNS tunneled traffic containing different protocols. The different average values of the bands hints that the average entropy of a series of tunneled network traffic can help distinguish the internally tunneled protocols. From this we set out to explore whether these trends and the mean entropy values can help us identify internally tunneled protocols.

## V. PROTOCOL PREDICTION EXPERIMENTS

Here, the dataset containing 20 test traffic captures and two ground truth captures are used to determine the similarity among the entropy distributions (variables) of plain network traffic of a particular protocol and its equivalent DNS tunneled versions. Intuitively, we use a simple similarity metric based on the averages of the distributions (variables). This metric was termed "MeanDiff", which is the shortened form of "Mean Differences". It is calculated simply as the absolute difference between the means of the two variables:

$$m(X, Y) = |\mu_X - \mu_Y|$$

The two variables being tested are: The *entropy* values of the ground truth protocol packet-capture (X) over time; and the *entropy* values of a given tunneled test capture (Y).

### A. Protocol Prediction Proof of Concept and Results

A small proof-of-concept program was built to evaluate the suitability of the chosen similarity metric for predicting the underlying protocol of a given DNS tunneled network traffic capture. The scripts used for the classifier are made available online in a Github repository online [23].

The proof of concept tool takes 2 ground truth traffic captures, each containing plain HTTP traffic and plain FTP traffic, respectively. It calculates the entropy values of each packet in the capture stream at the IP-packet level, generating 2 entropy distributions: one for HTTP traffic and the other FTP traffic. The tool then accepts a DNS tunneled traffic capture whose internally tunneled protocol is unknown. Random sampling is performed selecting a consecutive series of entropy values from the DNS tunneling capture. The random sample series length is set at 90% of the length of the given ground truth capture being compared against. 1000 samples are taken and the MeanDiff metric is calculated for each sample against the respective HTTP and FTP entropy distribution. An average of the 1000 rounds is then taken as the MeanDiff score against the respective ground truth entropy distributions (HTTP and FTP) respectively for a given DNS tunneled sample. This score is used as the basis for prediction, where the MeanDiff metric is taken as a distance metric. The ground truth protocol with the smallest MeanDiff score is deemed to be closer to the test DNS tunneled traffic sample. This ground truth protocol is predicted as the internally carried protocol.

This proof of concept tool was applied on the dataset and a sample run of the prediction results are shown in Table 1.

TABLE I. SAMPLE RUN RESULTS OF THE PROOF-OF-CONCEPT TOOL

| DNS Tunneled Test Sample | True Value | MeanDiff Prediction |
|---|---|---|
| 1[amazon] | HTTP | HTTP |
| 2[bbc] | HTTP | HTTP |
| 3[craigslist] | HTTP | FTP |
| 4[dsv.su.se] | HTTP | HTTP |
| 5[en.wikipedia] | HTTP | HTTP |
| 6[facebook] | HTTP | HTTP |
| 7[google] | HTTP | FTP |
| 8[youtube] | HTTP | FTP |
| 9[audio-wav] | FTP | HTTP |
| 10[audio-mp3] | FTP | HTTP |
| 11[img-jpg1] | FTP | FTP |
| 12[img-jpg2] | FTP | FTP |
| 13[img-png1] | FTP | FTP |
| 14[img-png2] | FTP | FTP |
| 15[pdf1] | FTP | FTP |
| 16[pdf2] | FTP | FTP |
| 17[txt1] | FTP | FTP |
| 18[txt2] | FTP | FTP |
| 19[video] | FTP | FTP |
| 20[zipfile] | FTP | FTP |
| **Percentage Correct (Accuracy):** | | **(15/20) = 75%** |

### B. Discussion

The MeanDiff metric in Table 1 shows a prediction *accuracy* of approximately 75%. Subsequent runs have shown results within the range of 70-80 percent prediction accuracy, showing promise for protocol prediction within DNS tunneled traffic. Here we also see that 5 out of the 8 HTTP over DNS test samples were classified correctly indicating a 62.5% *recall* (true-positive) rate, while 10 out of 12 FTP over DNS test samples were classified correctly indicating an 83.3% *recall*. The performance of the classifier is summarized in the confusion matrix in Table II.

TABLE II. PoC CLASSIFIER PERFORMANCE CONFUSION MATRIX

| N=20 | | Predicted | |
|---|---|---|---|
| | | HTTP | FTP |
| Actual | HTTP | 5 | 3 |
| | FTP | 2 | 10 |

The *Misclassification Rate* is seen to be 25%. The *Precision* for FTP is seen to be 76.9%, while for HTTP it is seen to be 71.4%. The *False Positive Rate* for the FTP class is seen to be 16.7%, while for the HTTP class it is seen to be 37.5%. These measures of the effectiveness of the classifier indicate positive results for the prediction of underlying network protocols in DNS tunnels.

## VI. CONCLUSIONS AND FUTURE WORK

In this study we considered the challenge of predicting the underlying application protocols tunneled within DNS traffic. The internal structure of DNS tunneling techniques was explored leading to the identification of a theory that the entropy distributions of the packet bytes can help characterize and predict the internally tunneled protocol. Packet traces were visualized in order to identify patterns emanating from different types of packets due to their content and function. A small dataset of DNS tunneled traffic was created, and proof of concept classifier was developed in order to test our theory. The classifier, based on the averages of the entropy values, showed a prediction accuracy of 75%, indicating a promising outlook for our theory.

Given the increasing use of DNS tunneling in security breaches this method aims to aid forensics practitioners to triage and identify DNS tunneling network traffic that may contain particular protocols of interest helping them focus on particular DNS tunnel flows and reducing the burden of having to analyze all DNS tunneled traffic.

From this study we provide initial results into the study of the relatively unexplored field of forensics on DNS tunneled network traffic. We provide a simple novel method of prediction of the internally tunneled network protocols within DNS tunneled traffic. Though 75% prediction accuracy is good, further improvements can be made to enhance this. One way to improve this could be to involve more features for the classification task. We only focused on information entropy while other features of the tunneled network traffic such as packet lengths, inter arrival times and character n-grams may also be analyzed in future studies. A wider analysis into other DNS tunneling techniques and more candidate internally tunneled protocols (plain, or encrypted) could help identify these. Also, other statistical metrics may offer finer grained differentiation and should be explored. Finally, machine learning and data mining techniques can also be explored to see if they provide better classification. We are presently looking into this, focusing particularly on Dynamic Time Warping in the field of time series analysis in order to find new ways to match the plain ground truth protocols against their DNS tunneled variants.